\documentclass[aps,prl,reprint,superscriptaddress]{revtex4-1}
\usepackage{graphicx}
\usepackage{rotating}
\usepackage{amssymb}    
\usepackage{amsmath}   
\usepackage{epsfig}
\usepackage[normalem]{ulem}   
\usepackage{bm}   
\usepackage{color}

\date{\today}

\begin{document}

\title{Evidence for crisis-induced intermittency during geomagnetic superchron transitions}

\author{Breno Raphaldini}
\email{brenorfs@gmail.com }
\affiliation{Instituto de Astronomia Geof\'isica e Ci\^encias Atmosf\'ericas, Universidade de S\~ao Paulo, 05508-090 S\~ao Paulo, Brazil}

\author{David Ciro}
\affiliation{Instituto de Astronomia Geof\'isica e Ci\^encias Atmosf\'ericas, Universidade de S\~ao Paulo, 05508-090 S\~ao Paulo, Brazil}

\author{Everton S. Medeiros}
\affiliation{Institute of Physics, University of S\~ao Paulo, Rua do Mat\~ao, Travessa R 187, 05508-090, S\~ao Paulo, Brazil}

\author{Lucas Massaroppe}
\affiliation{Instituto de Astronomia Geof\'isica e Ci\^encias Atmosf\'ericas, Universidade de S\~ao Paulo, 05508-090 S\~ao Paulo, Brazil}

\author{Ricardo Ivan Ferreira Trindade}
\affiliation{Instituto de Astronomia Geof\'isica e Ci\^encias Atmosf\'ericas, Universidade de S\~ao Paulo, 05508-090 S\~ao Paulo, Brazil}

\date{\today}

\begin{abstract}
 The geomagnetic field's dipole undergoes polarity reversals in irregular time intervals. Particularly long periods (of the order of $10^7$yrs) without reversals, named superchrons, have occurred at least three times in history. We provide observational evidence for high non-Gaussianity in the vicinity of a transition to and from a geomagnetic superchron, consisting of a sharp increase in high-order moments (skewness and kurtosis) of the dipole's distribution. Such increase in the moments is a universal feature of crisis-induced intermittency in low-dimensional dynamical systems undergoing global bifurcations. This suggests temporal variation of the underlying parameters of the physical system. Through a low dimensional system that models the geomagnetic reversals we show that the increase in the high-order moments during transitions to geomagnetic superchrons is caused by the progressive destruction of global periodic orbits exhibiting both polarities as the system approaches a merging bifurcation.  We argue that the non-gaussianity in this system is caused by the redistribution of the attractor around local cycles as global ones are destroyed.
 \end{abstract}

\maketitle

\section{Introduction}

The geomagnetic field is generated by a dynamo process taking place in Earth's outer core and is characterized by a strong dominant axial dipole, its polarity reverses in irregular time intervals that range from orders of $\sim 10^4$ to $5 \cdot 10^7$ years. Longer periods of order of $10^7$ years are named geomagnetic superchrons and have occurred at least three times in history, the Cretaceous Normal Superchron from about 120 to 83 million years ago, the Kiaman Reverse Superchron lasted from around 312 to 262 million years ago, and the Moyero Reverse Superchron, from 485 to 463 million years ago. In the literature, there is no current consensus concerning the cause of such long periods without geomagnetic reversals, some authors argue that these events may be generated by stationary chaotic and/or stochastic processes, implying that these events do not depend on eventual variations on physical parameters of the system \cite{Ryan2007}. On the other hand, other authors argue that superchrons are caused by non-stationary processes, i.e. long-term variations of natural parameters that determine the dynamics of Earth's core, such as the evolution of the heat flux patterns and other physical quantities that characterize the configuration of the core-mantle boundary \cite{Driscoll2011,Petrelis2011}. Moreover, the question of whether there exists a paleomagnetic warning preceding a superchron is a long-standing question in the field of geomagnetism \cite{Hulot2003}.

An agreement for this open debate relies on the paleomagnetic measurements of field directionality and intensity. Previous studies on this data have already provided important aspects of superchrons, for example, the existence of more stable geomagnetic field configuration during the superchrons \cite{Veikkolainen2014} and low field intensities just prior to a superchron transition \cite{Zhu2003, Zhu2001}. Additionally, some studies on geomagnetic dipole intensity indicate the existence of long-term trends, and possibly higher average values during superchrons \cite{Driscoll2016}. Nevertheless, the causes of the observed absence of field reversals remain elusive.

In order to address these questions we characterize the paleointensity measurements by estimating the high-order moments of their statistical distribution. This particular approach is motivated by the ability of distribution moments in detecting non-stationary transitions, tipping points, in real-world \cite{Gsell2016} and theoretical \cite{Guttal2008,Kefi2014} dynamical systems. Additionally, the high-order distribution moments have been used to investigate the departure from Gaussianity in ensembles of intrinsically chaotic solutions. This is particularly useful for characterizing the level of uncertainties in these systems \cite{Abramov2004}.

In this work we consider the third and fourth-order distribution moments, namely kurtosis and skewness, for time intervals of the paleointensity data of the geomagnetic field. We find that the transitions from and to superchrons are associated with a sharp increase of these statistical indicators near the begining and ending of the superchron, which strongly supports the hypothesis of non-stationary and bifurcation-related mechanisms for this scenario. Additionally, we show that both statistical moments, skewness and kurtosis, also have a sharp increase prior to a non-stationary transition from a reversing to a non-reserving state in a low order dynamical model for dipole intensity. Such a relation between low-dimensional dynamical systems and the infinite-dimensional hydromagnetic problem is supported by estimates on the dimension of the attractor from paleointensity data, which indicates a low dimensional intrinsic dynamics of the geomagnetic axial dipole \cite{Ryan2008}.

Finally, we demonstrate that the variations of the high-order moments in the low-dimensional model are associated with the systematic destruction of global unstable periodic orbits preceding the global bifurcation where the field reversal ceases. Here, the sequential destruction of cycles leads to the redistribution of the flow along the chaotic attractor, under the control of previously less influential cycles, causing in turn the observed changes in the distribution function momenta.

\section{Observational evidence for critical transitions leading to Superchrons}

Pint Database is a compilation of paleomagnetic measurements by various authors corresponding to the last 3 billion years \cite{Biggin2014}. In order to characterize the geomagnetic field during transitions to superchrons, we analyze the third (skewness) and fourth (kurtosis) moments of statistical distributions of the geomagnetic field intensity. These high-order distribution moments have been already associated with the proximity to non-stationary transitions \cite{Dakos2012}. Despite of eventual difficulties in precisely defining the exact location of transitions for oscillating trajectories \cite{Medeiros2017,Bathiany2018}, they proved to be useful in different contexts \cite{Guttal2008,Kefi2014,Gsell2016,Alam2013,Faranda2014,Fuentes2019}. 

\begin{figure}[h]
  \includegraphics[width=7cm,height=8cm]{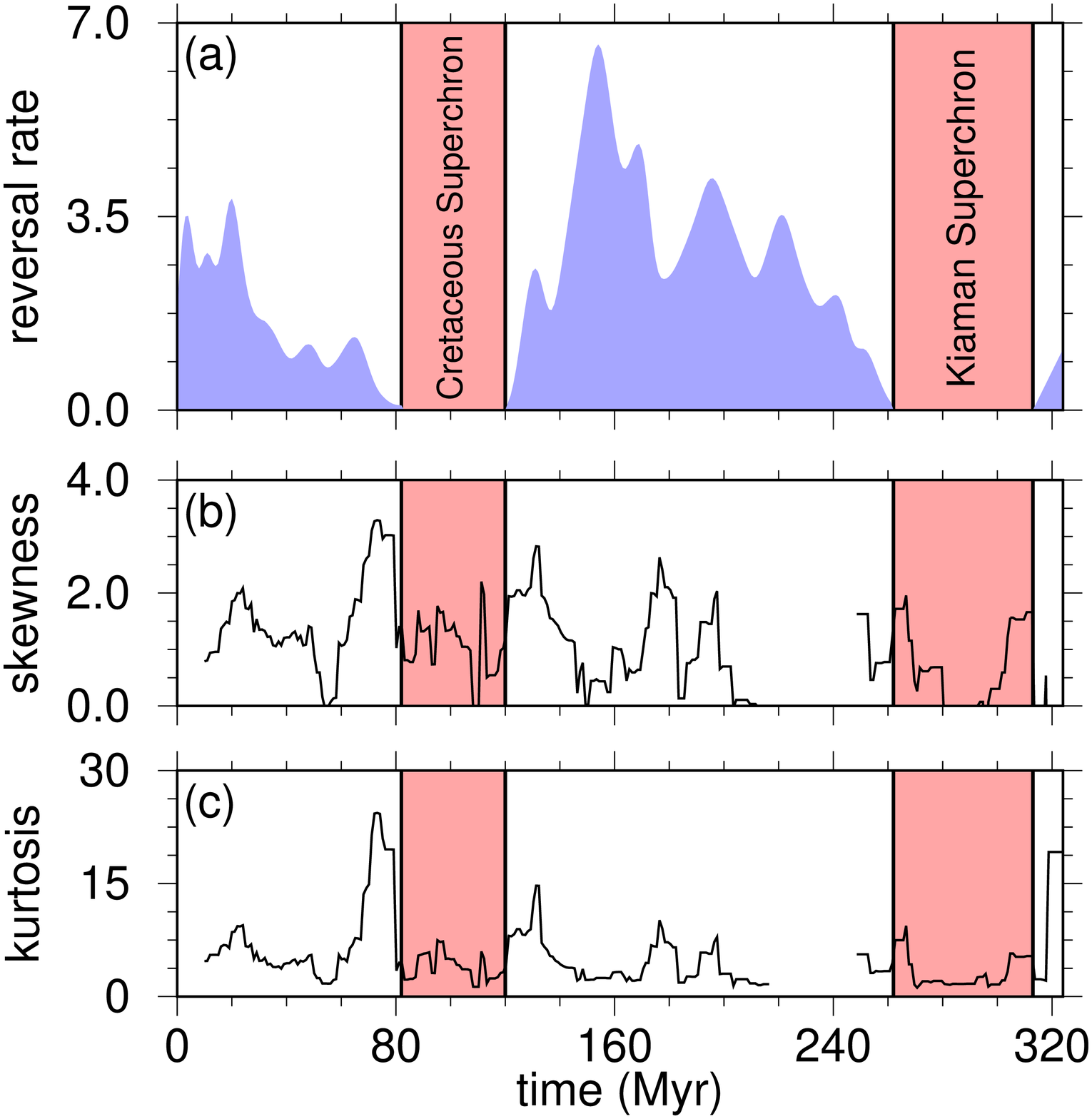}
  \caption{(a) Geomagnetic reversal rates for the Pint database. The time intervals marked in red indicate correspond to the superchrons. (b) The skewness obtained for a slinding window of size $\delta=2$ million years. (c) The kurtosis obtained for the same slinding window of (b). Notice the beginning and the end of a superchron episode is accompanied by a peak of the skewness and kurtosis.}
  \label{Figure1}
\end{figure}

Now, in the paleomagnetic measurements, we assume that the parameters that determine the geodynamo evolution are approximately constant in a timescale of a few million years, a reasonable assumption given that the timescale for the convection of Earth's mantle is at least one order of magnitude greater \cite{Richards1992}. Then, we calculate the moments at a given moment $T$, by averaging over all data in a time period $[T-\delta, T+\delta]$. Hence, in Figure~\ref{Figure1}(b) and Figure~\ref{Figure1}(c), we respectively show the skewness and the kurtosis for $\delta=2$ million years. The conspicuous peaks in both skewness and kurtosis appearing before and after both superchrons shown in Figure~\ref{Figure1}(a) are compelling evidence of the non-stationarity of these events. The applied statistical analysis is stable under the variation of $\delta$, see supplemental material for alternative values of $\delta$.

\begin{figure}[h]
  \includegraphics[width=1.0\linewidth]{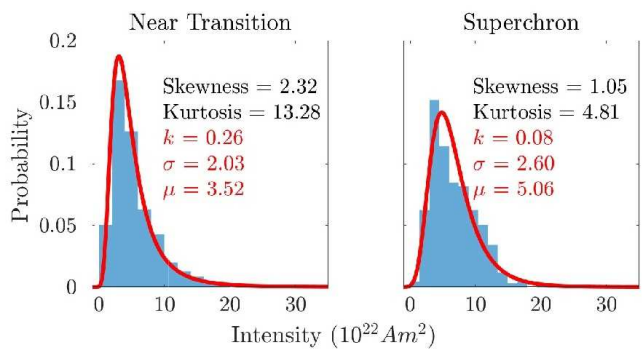}
  \caption{Histogram of the paleointensities in two different geodynamo regimes: low reversal rates near the transition to a superchron (middle) and superchron state (bottom) accompanied with the corresponding values of kurtosis and skewness. In each of the plots a generalized extreme value distribution probability density function was fitted to the data (red curve). The drop in the shape parameter k of the GEV distribution may me indicative of the transition.}
  \label{Figure2}
\end{figure}

The variations in the distribution moments observed in Figure~\ref{Figure1} indicate that during the transition from a normal reversing geodynamo state to a superchron, or vice-versa, the statistical distribution is fat-tailed and skewed to the left. These measurements deviate significantly from the Gaussian distribution ($\mbox{kurtosis}=3$ and $\mbox{skewness}=0$), implying that such transitions are preceeded by a higher frequency of extreme events in the geomagnetic field dipole intensity. Now, to further illustrate the fat-tailedness and the asymmetry of the intensity distribution near the transition, in Figure~\ref{Figure2} we obtain histograms of paleointensities in two regimes:  ii) near to the superchron transition, the frequency of extreme events in the field intensity increases (higher kurtosis) and its distribution become asymmetrical (higher skewness). i) During the superchron (no reversals), the field intensity is known to be more stable, this fact is reflected in our analysis by a relatively low value of kurtosis and skewness. Besides the moment analysis we have fitted the a generalized extreme value (GEV) distribution to the data, we show that the shape parameter k of the GEV distribution drops to low values in the superchron, which may be regarded as another indicative of a critical transition as shown by \cite{faranda2012generalized}.

\section{Low dimension dynamical system analysis}

In order to interpret the sharp increase in high order moments during a regime transition in the geodynamo, we use a low order model for geomagnetic field reversals consisting of a set of three nonlinearly coupled ordinary differential equations introduced in \cite{Gissinger2010} and further analyzed in \cite{Gissinger2012}. This model was derived directly from the induction equation by a low order truncation combined with symmetry arguments. Additionally, similar low dimensional equations can be derived for different truncation arguments \cite{Raphaldini2015}. Although this kind of approximations of the infinite dimensional magnetohydrodynamic equations may apear too drastic, this system keeps relevant qualitative features of the observed geodynamo \cite{Gissinger2012}. This argument is additionally supported by estimates on the dimension of the attractor from paleointensity data indicating a low dimensional nature of the geomagnetic axial dipole dynamics \cite{Ryan2008}. A more detailed discussion on the approximation of dissipative continuous systems, like the MHD equations governing the geodynamo, by appropriate three dimensional dynamical systems can be found in \cite{Robinson1998}.

In particular, the three-dimensional system of differential equations involving the magnetic dipole and quadrupole components $(Q,D)$, and a velocity field mode $(V)$, take the form:

\begin{eqnarray}
\nonumber
\dot{Q}&=&\mu Q-VD\\  
\label{model}
\dot{D}&=&-\nu D+VQ\\
\nonumber
\dot{V}&=&\Gamma - V+QD
\end{eqnarray}
where the parameters $\{\Gamma, \mu, \nu\}$ control the forcing intensity, the energy dissipation rate, and an instability growth rate, see \cite{Gissinger2010, Gissinger2012}. One of the advantages of this reduced model of geomagnetic reversals over others is the ability of reproducing the transitions from superchrons to normal reversing dynamo state by varying one parameter of the system as reported in \cite{Gissinger2012}. This is a non-stationary transition well characterized in bifurcation theory as a crisis-induced intermittency \cite{Grebogi1987}. Additionally, early warnings for this kind of global bifurcation involving chaotic attractors have been recently addressed in the literature \cite{Karnatak2017}. 

\begin{figure}[h]
  \includegraphics[width=7cm,height=8cm]{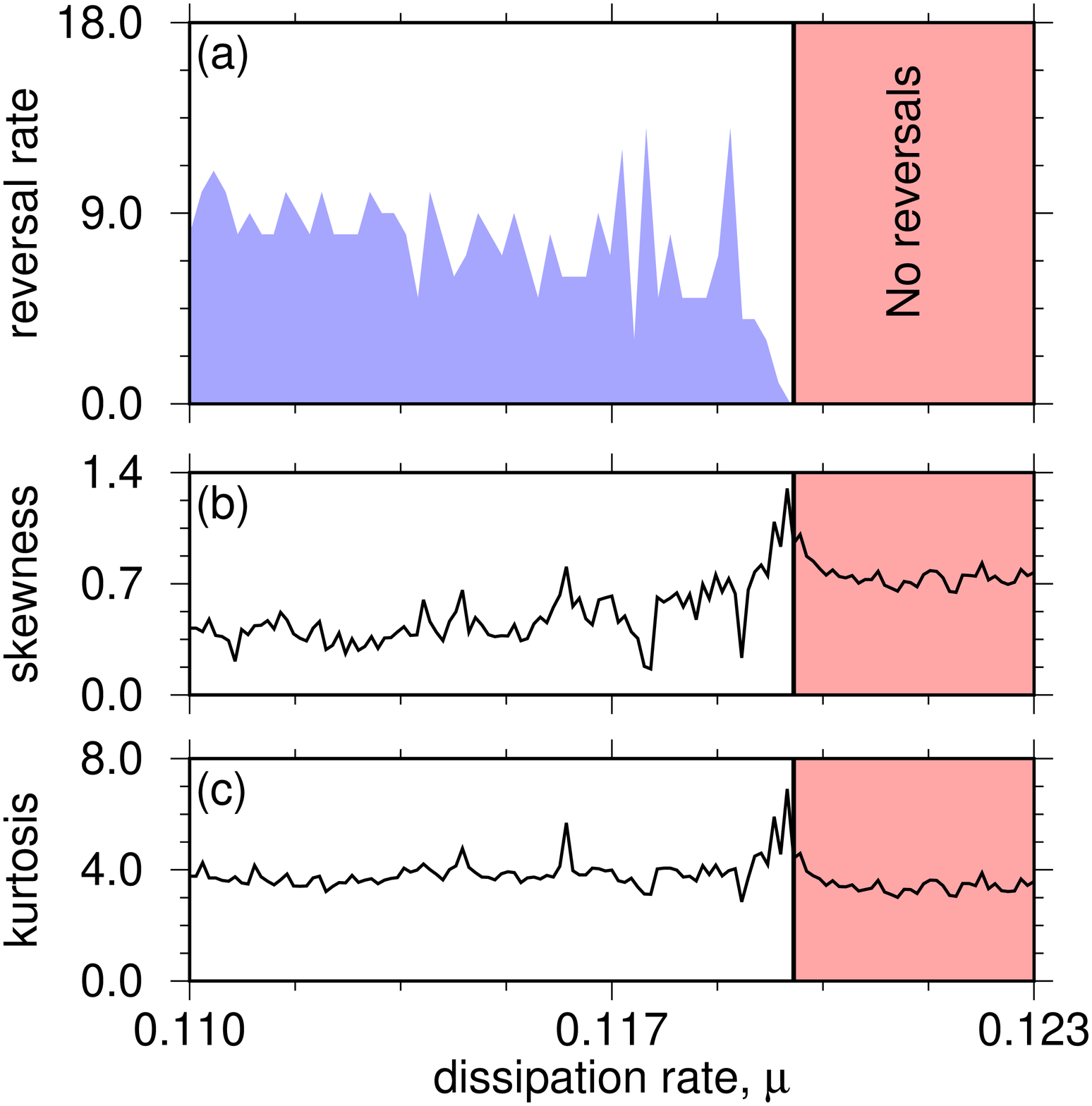}
  \caption{(a) The reversal rates obtained for $t=1000$ (arb. units) and for a range of dissipation rates of the model described by Eq.~(\ref{model}). (b) The corresponding skewness and (c) the kurtosis. Analogously to the observational results of paleomagnetic measurements, both kurtosis and skewness peak when the transition occurs.}
  \label{Figure3}
\end{figure}

Here, we aim to establish a connection between the statistical results obtained for the paleomagnetic measurements of the geomagnetic field with the non-stationary crisis-induced intermittency occurring in the model described by Eq.~(\ref{model}). For that, we investigate the behavior of the model during the crisis-induced intermittency an inverse merging bifurcation, which corresponds to the transition to a superchron in the paleomagnetic measurements. Hence, in Figure~\ref{Figure3}(a), the aforementioned model was integrated numerically for $200$ different values of the dissipation rate $\mu$. For every value of $\mu$, we integrate the system for $t=1000$ (arb. units) and compute the model reversal rate during this time interval. The reversal rate corresponds to the number of inversions performed by the dipole intensity of the system between two symmetric portions of a chaotic attractor. Next, for every value of $\mu$, we calculate the kurtosis and skewness of the distribution generated by the model dipole intensity during a time interval of $t=1000$ (arb. units). Notice that, in Figure~\ref{Figure3}(b) and Figure~\ref{Figure3}(c), there is a sharp increase in both moments when $\mu$ reaches a critical value $\mu_c$, at which the transition from a reserving state to a non-reversing, or superchron, occurs. 

Clearly, the behavior of the statistical moments across the critical value observed in Figure~\ref{Figure3} resemble the previously described changes in the third and fourth moments on the paleointensity measurements. These observations suggest the existence of a common mechanism between the reduced model and the geodynamo. Another common feature between the geodynamo and the reduzed model is that there can be increases in the third and fourth moments of the dipole intensity away from the critical time or parameters respectively, but such peaks apear for shorter periods of time or small parametric windows, and will be linked to another type of local bifurcation in the next section.

\section{Statistical moments and the destruction of global UPO's}

To better elucidate the mechanisms behind the increase in the third and fourth statistical moments we need to investigate the structure of the phase space before and after a transition. For the values of the parameters considered here, the system has three fixed points, being one in the $V$ axis labeled $C$, and two symmetric ones with positive and negative values of $D$, which well shall label $D_{-}$ and $D_{+}$. All the fixed points are saddles, with an one-dimensional stable manifold and a two-dimensional unstable manifold. In this model, a transition of the orbit from the vicinity of $D_{+}$ to the vicinity of $D_{-}$ constitutes a reversal of the system and the saddle $C$ only diverts orbits that come from far locations towards $D_+$ or $D_-$. In the non-reversing ``superchron'' state the system has two separated attractors, comprising the positive and negative $D$ states.

\begin{figure}[h]
  \includegraphics[width=0.8\linewidth]{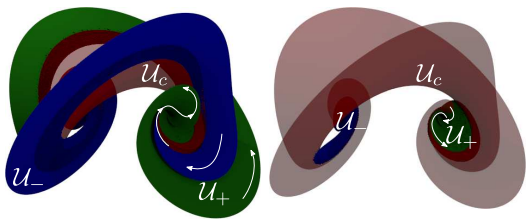}
  \caption{Unstable manifolds of $D_-$ (blue), $D_+$ (green) and $C$ (red) in a reversing case $ \mu < \mu_c$ (left), and non-reversing case $\mu>\mu_c$ (right). In the reversing case the unstable manifolds of $D_+$ and $D_-$ become inter-winded resulting in orbits that transit both regions of phase space, in contrast to the non-reversing case where the manifolds are confined on each side, resulting in orbits that do not reverse polarity. In both situations the unstable manifold of $C$ winds about both sides of the attractor diverting orbits coming from far in phase space.}
  \label{Figure4}
\end{figure}

\begin{figure*}[!htp] 
  \includegraphics[width=\linewidth]{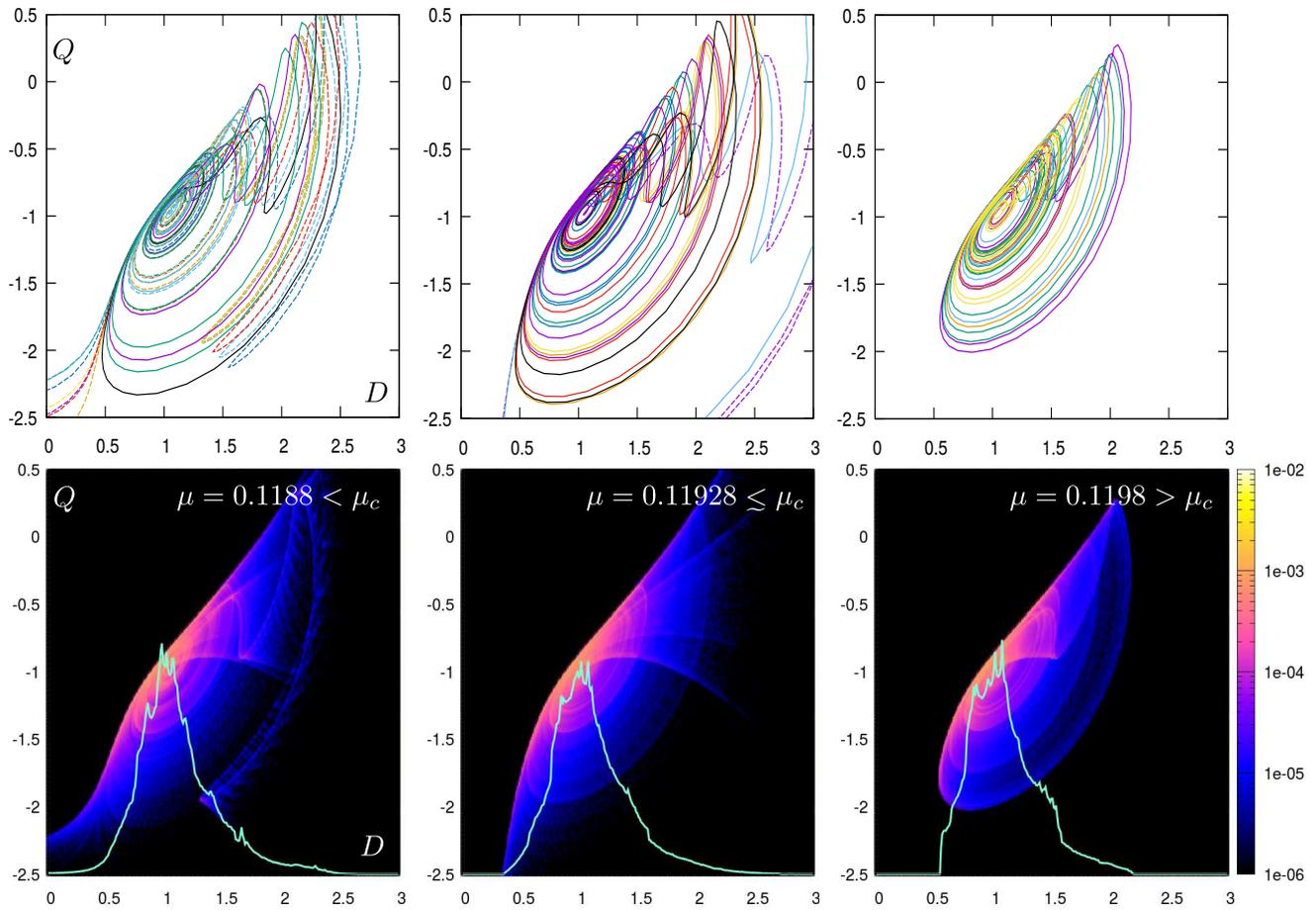}
  \caption{A few dominant UPOs in the $D-Q$ plane (top), for $\mu<\mu_c$ (left), $\mu\lesssim\mu_c$ (center) and $\mu>\mu_c$ (right), and associated 2D histograms with their respective 1D projection in the dipoles space. Local UPOs (continuous lines) are mostly concentrated near $D_{\pm}$, while global UPOs (dashed) are responsible for transitions. Near the critical value $\mu_c$ in a reversing state (center) new wider local and global UPOs generate extreme events, giving a fatter tailed distribution, when compared to $\mu$ far below or above $\mu_c$.}
  \label{Figure5}
\end{figure*}

In general, chaotic orbits are driven far from the saddles by the unstable manifolds $W^u(D_+,D_-)$ and attracted back by their one-dimensional stable manifold. This simple process generates an intricate set of closed orbits, or unstable periodic orbits (UPOs), which are periodic solutions of the differential equations, and have associated two-dimensional stable and unstable manifolds, in a similar fashion to the mentioned saddle points (Figure~\ref{Figure5}). Chaotic orbits bounce back and forth between UPOs, getting attracted through their stable manifold and then repealed through the unstable one, spending different amounts of time near each UPOs depending on its relative distance to the corresponding stable manifold. This process causes the UPOs to have an important influence on the phase-space distribution function, since orbits tend to spend long times in their vicinity.

A first description of the phase space structure can be made in terms of the manifolds of the saddles $D_+$, $D_-$ and $C$, which strongly influence the shape and extension of the UPOs. In the reversing state ($\mu<\mu_c$), both unstable manifolds $W^u (D_{+})$ and $W^u (D_{-})$ occupy a large region of the phase space and strongly interact by wrapping about each other in a complex helical spiral, as depicted in Figure~\ref{Figure4}-left, while the non-reversing situation ($\mu>\mu_c$) (Figure~\ref{Figure4}-right) the unstable manifolds $W^u(D_+)$ and $W^u(D_-)$ do not interact and are compactly wrapped about their corresponding saddles, separating the chaotic attractor in two pieces with different basins of attraction.

For a more detailed account of the phase-space distribution function, and, in particular, the dipoles distribution, we turn to the study of UPOs of the dynamical system, which can be created and annihilated as the control parameters are changed.
Transitional chaotic orbits are influenced by ``global'' UPOs, which access the vicinity of both antipodal fixed points (Figure~\ref{Figure5}-left), and only exist below the critical value $\mu_c$. For dissipation rates $\mu$ \emph{far} below $\mu_c$, the local and global UPOs are concentrated around $D+$ and $D_-$, leading to smaller values of the kurtosis and skewness. This changes as one approaches the critical value $\mu_c$ from below, global UPOs are sequentially destroyed, and new local and global UPOs wider in the dipole direction, with shorter life-span in $\mu$, are created in their place.

Since chaotic orbits tend to spend long times near the UPOs, each influential cycle causes an increase in the phase space distribution function around it. When changing parameters, some UPOs can be destroyed by touching some stable manifold of another invariant structure, spreading this higher density to other regions of phase space. Such spreading leads to an increase in the distribution moments respect to the previous situation with the underlying UPO.

This mechanism is quite general and explains the appearance of local peaks in the distribution moments away from the critical values in Figs.~\ref{Figure2},~\ref{Figure3}. Provided that all global UPOs must be destroyed before the critical value $\mu_c$, this redistribution process occurs very frequently near the global bifurcation, leading to the increase in the global high-order distribution moments for a longer time or a larger parameter window. Additionally, short-lived global cycles created during transition are wider than their predecessors, contributing further to the spreading of the distribution function during transition. Similar situation in which external UPO, that are usually invisible to the dynamics and become accessible though external forcing is described in other systems with geophysical background as in atmospheric dynamics, see \cite{gritsun2017fluctuations}, see also \cite{gritsun2013statistical} for further discussion on the role of the UPOs in the statistics of the atmospheric equations.

Wide UPOs also increase the occurrence extreme events in the magnetic dipole $D$, leading to a broader and more skewed distribution function (Figure~\ref{Figure5}-center). Local wide UPOs persist across $\mu_c$, but global ones cannot exist above it. Thus, the wider distribution function is maintained by wide local UPOs until they get destroyed as we further increase $\mu$, which is consistent with the subsequent decrease in the distribution moments of $D$, for $\mu>\mu_c$ as observed in Figure~\ref{Figure3}. Correspondingly, the local UPOs are compact and uniformly distributed on the attractor (\ref{Figure3}-right), in agreement with the small values of the kurtosis and skewness, which are maintained from here on.

\section{Concluding Remarks}

We have presented evidence for the occurrence of critical transitions just before and just after geomagnetic superchrons based on the sharp increase of the kurtosis and skewness of paleointensity geomagnetic data. This behavior is shown to be analogous to the observed in a low order model for geomagnetic reversals in which crisis-induced intermittency causes a shift of the system from a reversing state to a non-reversing one and presents the measured changes in the distribution moments of paleomagnetic data. This strongly suggests that the underlying mechanism that generates the transition in the Superchron geomagnetic transitions is driven by parametric changes instead of particularly long standings in a given polarity of a chaotic orbit. We argue that the mechanism proposed here to explain the sharp variation in the statistical moments is not strongly dependent on the particular low-dimensional model under consideration. Since type-one crisis-induced intermittency is defined by the merging of two (or more) separated attractors into one, it involves the interaction between invariant manifolds of different periodic orbits and fixed points. The sequential destruction of global UPOs near the transition leads to the redistribution of the higher distribution values around them, increasing the global kurtosis and also generating wider short-lived periodic orbits which further increase this distribution moment and are asymmetric respect to the antipodal points leading to a higher skewness. This moments increase also evidences the occurrence of more extreme events in the magnetic dipole intensity just before and after a transition to a superchron state.

\section{Acknowledgements}
This work was supported by the Brazilian agencies CAPES and FAPESP (Processes: 2017/23417-5, 2013/26598-0).

\section{Appendix A: Searching and tracking UPO's}

Consider a dynamical system in the autonomous form
\begin{equation}\label{eq.dyn_sys}
    \frac{dx}{dt} = f(x),
\end{equation}
where $x\in\mathbb{R}^n$ and $f:\mathbb{R}^n\rightarrow\mathbb{R}^n$. The solution of the equations defining the dynamical system is the flow $\varphi:\mathbb{R}^n\times\mathbb{R}\rightarrow\mathbb{R}^n$, of the vector field $f$. This flow is such that $\varphi(x,t)$, is the position of the solution starting at $x$ and evolving during a time $t$. The functional 
\begin{equation}\label{eq.functional}
    S(x,t) = |\varphi (x,t) - x|^2
\end{equation}
measures the distance between an initial condition $x$ and the flow departing from it after a time $t$. Clearly, if $x^*$ belongs to a periodic orbit of the dynamical system (\ref{eq.dyn_sys}), it satisfies $S(x^*,T)=0$, where $T>0$ is the time of first return, or period of the orbit.

Provided that we have the means to produce $\varphi(x,t)$, the numerical determination of UPOs consists in the identification of pairs $(x_i, T_i)$ for which the functional (\ref{eq.functional}) vanishes. Consider the collective variable $u\in\mathbb{R}^{n+1}$, such that $u_i = x_i$ for $i\in\{1,2,...,n\}$ and $u_{n+1} = t$. We want to minimize the functional
\begin{equation}
 S(u) = \sum_{i=0}^n [\varphi_i (u) - u_i]^2.
\end{equation}
Following the Levenberg-Marquardt procedure we start from $u^0$ which is close to the desired solution $u^*$, then consider a small variation $\delta u$, such that $u_0+\delta u = u^*$, causes $S(u^0+\delta u)$ to become stationary respect to $\delta u$. This leads to an approximated equation for $\delta u$
\begin{equation}\label{eq_lev-marq}
 [A^T A - \lambda\mbox{Diag}(A^T A)] \delta u = A^T [x - \varphi(u)]
\end{equation}
where $A_{i,j} = \partial\varphi_i/\partial u_j|_{u^0} - \delta_{i,j}$, and $i=1,2,...,n$, $j=1,2,...,n+1$. Diagonal damping is included to prevent overshooting and scale appropriately the steps on each direction. Since (\ref{eq_lev-marq}) is not exact, the procedure must be iterated until the functional goes below some tolerance value. Additionally, the scalar $\lambda$ is an adaptive parameter that allows to control the velocity of convergence if the method approaches stagnation.

An important issue is the numerical determination of $\partial\varphi_i/\partial u_j|_{u_0}$, which are the elements of the monodromy matrix when the solution is periodic. This matrix can be obtained by integrating
\begin{equation}\label{eq.monodromy}
 \frac{dA}{dt} = J(x(t)) A(x(t)),
\end{equation}
where $A(x(0))$ is the identity, $x(t)$ is a numerical solution of the system starting at $x(0)=x_0$, and $J$ is the regular Jacobian of the flow. Finally, for the time derivative of the flow we have simply
\begin{equation}
   \frac{\partial}{\partial t}\varphi(x,t) = f(\varphi(x,t), t).
\end{equation}
Provided that the reference orbit is sufficiently close to the UPO, equation (\ref{eq.monodromy}) should give a reasonable approximation of the monodromy matrix. An additional complication emerges near the UPO destruction, where the system (\ref{eq.monodromy}) becomes unstable and a special method of finite differences becomes more reliable and easier to implement.

\subsection{Determining initial conditions}
Provided that chaotic orbits are subsequently attracted to and repelled from different Unstable Periodic Orbits (UPOs), the chaotic dynamics contain ordered sequences of the underlying periodic orbits. This mimicking allow us to use chaotic orbits to estimate the locations of UPOs and their periods. In principle, we can take $N$ subsequent points that discretize a finite segment of the chaotic orbit and perform $N^2$ comparisons to see which points are separated in time and very close in distance. However, this can be extremely expensive and the following pre-selection process was implemented:

\begin{itemize}
    \item Integrate the dynamical system from an arbitrary $x^{\dagger}$ for a very long time. Define $\varphi(t) = \varphi(x^{\dagger},t)$
    \item Determine the instants $\{t_{i}\}$ for which some apropriate component of the flow $\varphi_k(t_i)$ reaches a maximum value.
    \item Measure the distances $d_{i,j} = |\varphi(t_{i})-\varphi(t_{j})|$.
    \item If $d_{i,j}<\epsilon$ for some small $\epsilon$ and $j>i$, define the initial guess $u_0 = \{\bar\varphi_{i,j},\Delta t_{i,j}\}$, with $\bar\varphi_{i,j} = [\varphi(t_i) + \varphi(t_j)]/2$ and $\Delta t_{i,j} = t_j - t_i$. Take the new $i$ equal to $j+1$ to prevent initial guesses that lead to the same orbit.
\end{itemize}

This method uses the simple fact that periodic orbits must contain subsequent maxima and minima in all its dynamical variables. Then we can take the maximum of any dynamical variable as starting and ending point of the periodic orbit. If a chaotic orbit is mimicking this orbit it will develop two maxima in the same variable at approximately the same location, and the time between them will be a good initial guess of the period of the orbit. Additionally, by defining the initial guess using a reference chaotic orbit we ensure that the UPOs being tracked are the ones that really affect the most our dynamics, so that they contribute the most to the statistical moments of the distribution function.

\subsection{Tracking across parameters}
Provided that we have determined an UPO for a given value of $\mu$, we can take its period and \emph{any} point in the orbit as an initial guess for determining the UPO in $\mu+\delta_mu$. However UPOs can undergo important geometrical changes under small variations in the control parameters and even be destroyed, so, special care must be devoted to tracking these orbits in a sufficiently smooth fashion. To do this we take a relevant parameter of the orbit, e.g. its period $T(\mu)$, which must be an smooth function of the control parameter, and attempt to perform small fixed steps in such parameter instead of $\mu$, i.e. $T(\mu+\delta\mu) < T(\mu)+\delta T$, for some small fixed $\delta T$, then for small $\delta\mu$ we have
\begin{equation}
 \delta \mu \approx \delta T \left(\frac{dT}{d\mu}\right)^{-1}, 
\end{equation}
i.e. we can estimate $\mu_{n+1}$ from two close previous values
\begin{equation}
 \mu_{n+1} \approx \mu_n+\frac{\mu_n-\mu_{n-1}}{T_n-T_{n-1}}\delta T,
\end{equation}
where $T_n=T(\mu_n)$. For an appropriate choice of the period step $\delta T$ there is a better chance of converging to the appropriate UPO, in constrast to just taking fixed steps in $\mu$.

\end{document}